\documentstyle[12pt]{article}

\def\sqr#1#2{{\vcenter{\hrule height.#2pt\hbox{\vrule width.#2pt
height#1pt \kern#1pt \vrule width.#2pt}\hrule height.#2pt}}}
\def\square{\mathchoice\sqr64\sqr64\sqr{4.2}3\sqr{3.0}3}

\begin{document}

\def \be{\begin{equation}}
\def \ee{\end{equation}}
\def \ba{\begin{eqnarray}}
\def \ea{\end{eqnarray}}
\def \nn{\nonumber}

\begin{titlepage}
\title{ \LARGE Gauge Fixing in Higher Derivative Gravity 
\\
\author{ A.Bartoli \\ {\it Dipartimento di Fisica, Universit\`a di Bologna,} 
\\
{\it I.N.F.N. Sezione di Bologna, Via Irnerio 46, Bologna Italy.}\\
\\ J.Julve 
\thanks{ Partially supported by DGICYT and the 
CNR-CSIC cooperation agreement.} $\,$ and E.J.S\'anchez  \\ 
{\it Instituto de Matem\'aticas y F\'{\i}sica Fundamental,} \\
{\it CSIC, Serrano 123, Madrid, Espa\~na.}}}

\maketitle

\begin{abstract}
Linearized four-derivative gravity with a general gauge fixing term is considered. By a Legendre 
transform and a suitable diagonalization 
procedure it is cast into a second-order equivalent form where the nature of the physical degrees 
of freedom, the gauge ghosts, the Weyl ghosts, and the intriguing ``third ghosts", characteristic 
to higher-derivative theories, is made explicit. The symmetries of the theory and the structure 
of the compensating Faddeev-Popov ghost sector exhibit non-trivial peculiarities.
\end{abstract}

\vfill
Preprint IMAFF 98/09

\end{titlepage}

\section*{Introduction}
\parindent 1 cm

Theories of gravity with terms of any order in curvatures arise as part of the low energy 
effective theories of the strings [1] and from the dynamics of quantum fields in a curved 
spacetime background [2]. Theories of second order 
(four-derivative theories in the following) have been studied more closely in the literature 
because they are renormalizable [3] in four dimensions. This property spurred Renormalization--
Group studies [4-8], including attempts to get rid of the Weyl ghosts  (also known as 
``poltergeists") usually occurring in higher-derivative (HD) theories. On the practical side, HD 
gravity greatly affects the effective potential and phase transitions of scalar fields  in curved 
space-time, with a wealth of astrophysical and cosmological properties [9]. These 
phenomenological applications contributed to keep alive the theoretical interest, as illustrated 
by the most comprehensive introductory study available [10]. 

Besides the renormalization properties [3], all that was known about the structure of the 
(classical) theory was the particle contents, as read out of the linear decomposition of the HD 
propagator into pieces with second order (particle) poles. Some related aspects of the equations 
of motion [11] were also elucidated. Definite theoretical progress came from a procedure, based 
on the Legendre transformation, devised to recast four-derivative gravity as an equivalent theory 
of second differential order [12]. A suitable diagonalization of the resulting theory was found 
later [13] that yields the explicit independent fields for the degrees of freedom (DOF) involved 
(usually including massive Weyl ghosts), thus completing the order-reducing procedure. One should notice that theories with terms of higher order in curvatures have the same DOF and propagators of the four-derivative one, since the higher terms do no contribute to the linearized theory.

An alternative order-reducing method has been proposed [14] that introduces an auxiliary field coupled to the Einstein tensor $G_{\mu\nu}$ (or to the scalar curvature $R$) and featuring a squared term. It can be shown that this method is equivalent to the one based on a Legendre transformation with respect to $G_{\mu\nu}$ (or to $R$), the auxiliary field being a redefinition of the "momentum" conjugate to them.

The studies [12-14] above were carried out for the (non quantizable) 
Diff-invariant theory. An exploration of the method in the presence of gauge fixing terms has 
been done in a simplified HD gauge field model [15]. In this paper we implement this procedure 
for four-derivative gravity. 

Amongst a crowd of positive and negative norm, gauge-independent and gauge ghost, masless and 
massive states, the famous ``third ghosts" arise.  These subtle ghosts, missed in [4] and 
properly accounted for in [5] and there since, first emerged from a functional determinant in the 
context of Path Integral quantization. Now they appear as the poltergeist partners of the usual 
gauge ghosts.  

In Section 1 we present our starting Diff-invariant four-derivative theory.
A very general gauge fixing term is then introduced that includes the most used ones  as 
particular cases. Being interested in the propagators and
in the ensuing DOF identification, we focus mainly on the free part of the Lagrangian. Self-
interactions and interactions with other matter fields are
embodied in a source term and may be treated perturbatively. Then the relevant total gauge fixed 
linearized theory is worked out.  Section 2 deals with the order-reducing procedure that leads to 
the diagonalized second-derivative equivalent theory.
The structure of the propagators and the identification of the DOF is then worked out in Section 
3. The Faddeev--Popov (FP) compensating Lagrangian is studied in Section 4, where an order-
reduction of the fermion sector is also carried out. Particular attention is paid here to the  
identification of poles and to the striking cancellation mechanism of ghost loop contributions. 
Related to this, a discussion of the BRST symmetries involved is also made. The above results are 
summarized and further elucidated in the conclusion.

The definitions of the spin projectors and related formulae, the basis of local differential 
operators, and the notations and conventions used throughout the paper have been collected in 
Appendix I in order to render the work almost self-contained and more readable. Secondary 
calculations regarding the conditions from locality on the gauge-fixing parameters and the order-
reduction of the HD fermionic FP Lagrangian have been respectively moved to two Appendices.

\section{The Linearized Lagrangian}
We consider a general four-derivative theory of gravity 
\begin{equation}
{\cal L}_{HD} = {\cal L}_{inv} + {\cal L}_{g}+{\cal L}_{m}\quad ,
\end{equation}
where ${\cal L}_{m}$ is the coupling with matter,
\begin{equation}
{\cal L}_{inv} = \sqrt g \ [ aR + bR^2 + c R_{\mu \nu} R^{\mu\nu} ] 
\quad ,
\end{equation}            
is the most general Diff-invariant gravitational Lagrangian of second order in curvatures
(the squared Riemann tensor has not been considered as long as a topologically trivial 4D 
spacetime is assumed so that the Gauss-Bonnet identity holds), and
\begin{equation}
{\cal L}_{g} = \sqrt{g} \frac{1}{2} 
        \chi^{\mu}[h] {\cal G}_{\mu  \nu}\chi^{\nu}[h]\, ,
\end{equation}
where
\begin{eqnarray} 
 \chi^{\mu}[h] &\equiv& A^{\mu} - \lambda D^{\mu} h \quad , \\
{\cal G}_{\mu \nu} &\equiv& \xi_1 D^{\rho}D_{\rho} g_{\mu\nu} - 
\xi_2\frac{1}{2}D_{(\mu} D_{\nu)} + \xi_3 g_{\mu\nu} 
+\xi_4 R_{\mu \nu} + \xi_5 R g_{\mu \nu} \quad ,
\end{eqnarray}
is a general gauge fixing term  that depends on six gauge parameters and generally contains HD as 
well as lower-derivative (LD) terms. One may obtain the gauge fixings used in [4]-[8]  by 
specializing these parameters.
\par
 
In order to study the propagating DOF of the theory we work the quadratic 
terms in  $ h_{\mu \nu} $ out of $ {\cal L}_{HD} $. Dropping
total derivatives, they write  
\begin{equation}
{\cal L}_{HD}  =  {\cal L}^{(2)}_{inv} + {\cal L}^{(2)}_{g} +
{\cal L}_{s}
               = \frac{1}{2}  h^{\mu\nu} 
                 (P_{\mu \nu,\rho\sigma}^{inv} + P_{\mu \nu,\rho\sigma}^{g} ) 
                 h^{\rho\sigma} 
                 +{\cal L}_{s}\nonumber \quad .
\end{equation}
        The source term ${\cal L}_{s}$ includes the interactions with matter 
fields $\phi$ and all the self-interactions of $h_{\mu \nu}$ affected by the 
Newton constant $G_{N}$.
Here and in the following the indices are rised and lowered by 
$ \eta_{\mu \nu} $ and usually omitted for simplicity whenever no ambiguity 
arises.

The differential operator kernel for the diff-invariant part is 
\begin{equation}
P^{inv} = a \Box \left[ \frac{1}{2} P^{(2)} - P^{(S)} \right]
+ 6b \Box^2 P^{(S)} + c \Box^2 \left[ \frac{1}{2} P^{(2)} + 2P^{(S)} \right] \quad .
\end{equation}
The gauge fixing contribution
\begin{equation}
{\cal L}^{(2)}_{g} = \frac{1}{2}  ( A^{\mu} - \lambda \partial^{\mu} h ) 
	(\xi_1 \Box \eta_{\mu\nu} 
		- \xi_2\partial_{\mu} \partial_{\nu} + \xi_3 \eta_{\mu\nu} ) 
 ( A^{\nu} - \lambda \partial^{\nu} h ) 
\end{equation}
yields
\begin{eqnarray}
P^{g} & = & - \Box \lambda^2 \left[  ( \xi_1 - \xi_2 )\Box
+ \xi_3  \right] \left(3P^{(S)}+ P^{(W)}+ P^{\{SW\}}\right) \nonumber \\  
& & \mbox{} + \xi_2 \Box^2 P^{(W)} 
- \Box \left[\xi_1\Box +\xi_3  \right]
        \left(\frac{1}{2}P^{(1)} + P^{(W)}\right) \\
& & \mbox{} + \lambda \Box \left[ (\xi_1 - \xi_2)\Box + \xi_3 \right] 
        \left( 2P^{(W)} + P^{\{SW\}}\right) \nonumber\, .
\end{eqnarray}
One recognizes in (8) the linearized $\chi^{\mu}[h]$ and the $h$-independent part of 
$\cal{G}_{\mu\nu}\,$, which we call ${\cal G}^{(h)}$ in the following.

\noindent{Thus} the complete HD differential operator kernel is
\begin{eqnarray}
P  & = & P^{inv} + P^{g} \nonumber\\
&=& \frac{1}{2}\Box \left( c\Box+a \right) P^{(2)} 
        -\frac{1}{2}\Box \left( \xi_{1}\Box +\xi_{3}\right) P^{(1)}
                \nonumber 
\\
& & \mbox{} + \Box\left[ -a+2(3b+c)\Box -3 \lambda^{2}
        \left( (\xi_{1}-\xi_{2})\Box+\xi_{3} \right)
        \right] P^{(S)}
\\
& & \mbox{} -(\lambda -1)^{2} \Box\left( (\xi_{1}
        -\xi_{2} )\Box +\xi_{3}\right) P^{(W)}
                \nonumber 
\\
& & \mbox{} -\lambda (\lambda -1)\Box \left( (\xi_{1}-\xi_{2})\Box +\xi_{3}\right)
        P^{\{SW\}} \nonumber\, .
\end{eqnarray}
\noindent{B}y decomposing $ P $ in its HD and LD parts, namely
\begin{equation}
P = M\Box^2  + N\Box  \quad ,
\end{equation}
where 
\begin{eqnarray}
M &\equiv& \frac{c}{2} P^{(2)}- \frac{1}{2} \xi_1 P^{(1)} \nonumber \\
& & \mbox{} + \left( 2(3b+c)-3\lambda^{2}(\xi_{1}-\xi_{2})\right)P^{(S)}  
-(\lambda -1)^{2}(\xi_{1}-\xi_{2})P^{(W)}   \nonumber \\
& & \mbox{}
-\lambda (\lambda -1) (\xi_{1}-\xi_{2})P^{\{SW\}} \\
N & \equiv&  \frac{1}{2} a P^{(2)} 
        - \frac{1}{2} \xi_3  P^{(1)}\nonumber \\
        & & \mbox{} -\left(a +3\lambda^{2}\xi_{3} \right)P^{(S)}
        -(\lambda -1)^{2}\xi_{3}P^{(W)}
        -\lambda (\lambda -1)\xi_{3} P^{\{SW\}}\nonumber \quad ,
\end{eqnarray}
equation (6) may be written as 
\begin{equation}
{\cal L}_{HD} = \frac{1}{2} h \Box (M\Box  + N) h 
+{\cal L}_{s}\quad .
\end{equation}
Dropping total derivatives, it can be given the more 
convenient form
\begin{equation}
{\cal L}_{HD} [h,\Box h]
               = \frac{1}{2}(\Box h)M(\Box h) + \frac{1}{2}hN(\Box h) 
               +{\cal L}_{s}\quad .
\end{equation}
The HD  Euler's equation takes also the form
\begin{equation}
\Box (M\Box  + N)^{\mu \nu ,\rho \sigma} h_{\rho \sigma} = T^{\mu \nu}\quad ,
\end{equation}
where $T^{\mu \nu}\equiv-\frac{\delta {\cal L}_{s}}{\delta h_{\mu \nu}}\;$.

\section{Second order equivalent theory}

In order to perform a Lorentz-covariant Legendre transformation [13][16]
of our HD Lagrangian, the form of (14) trivially suggests defining the 
conjugate variable
\begin{equation}
\pi^{\mu\nu} =  \frac{\partial {\cal L}_{HD}}{\partial (\Box h_{\mu\nu})}\quad .
\end{equation}
One finds 
\begin{equation}
\pi = M (\Box h) + \frac{1}{2} Nh  +O(G_{N})\quad ,
\end{equation}
where the contributions from the gravitational interactions may be 
accounted for perturbatively in $G_{N}\,$, or may be simply ignored for
the analysis of the propagating DOF.

As required, (17) is invertible and gives
\begin{equation}
\Box h = M^{-1} \left[\pi - \frac{1}{2}Nh \right] \equiv F[  h , \pi ] 
\quad .
\end{equation}
Notice that the operators $ M $ and $ N $ are invertible as long 
as gauge fixing terms have been introduced. Otherwise they would project into 
the spin-state subspace $ 2 \oplus S $, then being singular.\par
The Lorentz-covariant Hamiltonian-like function is then 
\begin{eqnarray}
{\cal H} [ h , \pi ] & = & \pi F[ h,\pi ] 
                - {\cal L}_{HD}[ h ,F[ h ,\pi ]] \nonumber \\
& = & \frac{1}{2} \left[\frac{1}{2}N h - \pi\right]
 M^{-1} \left[ \frac{1}{2}N h - \pi\right]-{\cal L}_{s}\quad .
\end{eqnarray}
The  equations of motion turn out to be the system of canonical-like equations 
\begin{eqnarray}
  \Box h & = & \frac{ \partial {\cal H} }{ \partial \pi}  \\
\Box \pi & = &  \frac{ \partial {\cal H} }{ \partial h }
\; .
\end{eqnarray} 
The familiar negative sign one woud expect in (21) is 
absent because the definition (16) involves second derivatives
of the field $h$ instead of the usual velocity [15].
They may also be derived by a variational principle from the so called 
(now second-derivative) Helmholtz Lagrangian 
\begin{equation} 
{\cal L}_{H} [ h, \pi] = \pi \Box h - {\cal H} [ h, \pi ]\quad .
\end{equation}
In fact from (22) one sees that (20) is the Euler's equation for $ \pi $ and 
(21) is the one for $ h $. From (20) (which is nothing but equation (18)) 
one may
work out $ \pi $ as given by (17). Substituting it in (21) one recovers (15),
namely the original HD equation of motion. 

Mixed $ \pi - h $ terms occur in (22). The diagonalization can be 
performed by 
defining new tilde fields such that 
\begin{eqnarray}
 h & = & \tilde h + \tilde\pi \nn \\
\pi & = & \frac{N}{2} (\tilde h - \tilde\pi ) \; .
\end{eqnarray}
or conversely 
\begin{eqnarray}
\tilde h & = &  N^{-1} \left[\frac{1}{2}Nh + \pi\right] \nn \\
\tilde \pi & = &  N^{-1} \left[ \frac{1}{2}Nh -\pi\right]\; .
\end{eqnarray}
Then $ {\cal L}_{H} $ finally becomes the desired LD theory
\begin{eqnarray}
{\cal L}_{LD} &=& \frac{1}{2} \tilde h  N \Box \tilde h - 
\frac{1}{2} \tilde \pi ( N\Box+ NM^{-1}N ) \tilde \pi
+{\cal L}_{s}\quad ,
\end{eqnarray}
where 
\begin{eqnarray}
NM^{-1}N&=&\frac{a^{2}}{2c}P^{(2)}-\frac{\xi_{3}^{2}}{2\xi_{1}}P^{(1)}
                                        \nonumber\\
       & & \mbox{} 
       + \frac{a^{2}(\xi_{1}-\xi_{2})-3\lambda^{2}\xi_{3}^{2}2(3b+c)}
        {2(3b+c)(\xi_{1}-\xi_{2})}P^{(S)}\nonumber\\
        & & \mbox{}  -\frac{(\lambda -1)^{2}\xi_{3}^{2}}
        {\xi_{1}-\xi_{2}}P^{(W)}
        \\
& & \mbox{} -\frac{\lambda (\lambda -1) \xi_{3}^{2}}{\xi_{1}-\xi_{2}}
                        P^{\{SW\}}\nonumber
\end{eqnarray}
For further discussion, a most enlightening expression for (25) is obtained by separating the 
gauge-dependent parts
\begin{eqnarray}
{\cal L}_{LD} &=& \frac{1}{2}\tilde{h} a \left(
                \frac{1}{2}P^{(2)}-P^{(S)}
                \right)\Box\tilde{h}
+\frac{1}{2} \chi[ \tilde{h} ] {\cal G}^ {\tilde{(h)}} \chi[\tilde{h}] \nn\\
&&\mbox{} -\frac{1}{2}\tilde{\pi}\left[ 
a\left( \frac{1}{2}P^{(2)}-P^{(S)} \right)\Box   
+\frac {a^{2}}{2c} P^{(2)} 
+ \frac{a^{2}}{2(3b+c)} P^{(S)} \right]\tilde{\pi}\\
&&\mbox{}
-\frac{1}{2} \chi[ \tilde{\pi} ] {\cal G}^{\tilde{(\pi)}} \chi[ \tilde{\pi}]
+{\cal L}_{s}
\nn
\end{eqnarray}
where
\begin{eqnarray}
{\cal G}_{\alpha \beta}^{(\tilde h)} 
&=&\xi_{3} \theta_{\alpha \beta} +\xi_{3} \omega_{\alpha \beta} 
=\xi_{3} \eta_{\alpha \beta }
\\
{\cal G}_{\alpha \beta}^{(\tilde \pi)}&=&\xi_{3} 
 \frac{\xi_{1}\Box+\xi_{3}}{\xi_{1}\Box}  \theta _{\alpha \beta} 
+\xi_{3} \frac{(\xi_{1}-\xi_{2})\Box+\xi_{3}}{(\xi_{1}-\xi_{2})\Box} 
\omega_{\alpha \beta}          \quad ,
\end{eqnarray}
and the form of $\chi$ has been displayed in (8).

The physical meaning is now apparent: 
$\tilde h$ and $\tilde \pi$ describe the massless and the massive DOF  of the theory 
respectively. Notice that the gauge-invariant part of the kinetic term of $\tilde \pi$ reproduces 
that of the Fierz-Pauli theory [17].

        The LD Lagrangian (27) thus obtained is non-local for arbitrary  
gauge parameters. However, we can have locality for a particular choice
of parameters (see Appendix II). Even for this choice, an unpleasant feature of (27) is that the 
scalar subspaces $S$ and $W$ appear mixed as long as the transfer operator $P^{\{SW\}}$ occurs in 
$N$ and $NM^{-1}N$.

\section{Linearized theory and propagators}

In order to avoid unessential complications due to the $S$-$W$ mixing  that obscures the 
identification of the propagating DOF, one may redefine the field $ h_{\mu\nu} $ as
\begin{equation}
\hat{h}_{\mu\nu} = ( Q^{-1})_{\mu\nu} ^{ \rho\sigma} \; h_{\rho \sigma}
\end{equation}
where
\begin{eqnarray}
Q(\lambda) &=&  P^{(2)} + P^{(1)} +\frac{2}{3}P^{(W)}
 - \frac{2}{9} \frac{(\lambda -1)}{\lambda } P^{\{SW\}} 
\end{eqnarray}
is invertible and becomes a numerical matrix for $\lambda =-2\,$,
namely $Q(-2)=\bar{\eta}-\frac{1}{3}\bar{\bar{\eta}}\,$. This choice is
compulsory if we wish to avoid polluting the source term
with non-locality. The operator $ P $ transforms to
\begin{equation}
\hat{P}=QPQ= {\hat M}\Box^2 +  {\hat N}\Box \quad , 
\end{equation}
where
\begin{eqnarray}
\hat {M} & \equiv& \frac{c}{2} P^{(2)}- \frac{1}{2} \xi_1 P^{(1)}   \nonumber\\
& & \mbox{} + \frac{4}{27} \frac{(\lambda -1)^2}{\lambda^2} 2(3b + c)P^{(W)}
     - \frac{4}{27} \frac{(\lambda -1)^4}{\lambda^2}(\xi_1 - \xi_2)P^{(S)}  \quad ,
      \\
\hat{N} & \equiv&   \frac{1}{2} a P^{(2)}       
        -\frac{4}{27} a \frac{(\lambda -1)^2}{\lambda^2} P^{(W)} 
         - \frac{1}{2} \xi_3  P^{(1)}  
       - \frac{4}{27} \frac{(\lambda -1)^4}{\lambda^2}\xi_3  P^{(S)}  
       \nn
\end{eqnarray}
do not contain the operator $P^{\{SW\}}$ anylonger. Then equation (13) may be written as 
\begin{equation}
{\cal L}_{HD} = \frac{1}{2} \hat{h} \Box ( \hat{M}\Box + 
\hat{N}) \hat{h} +{\cal L}_{s}\quad ,
\end{equation}
or, dropping total derivatives,
\begin{equation}
{\cal L}^{(2)}_{HD} [\hat{h},\Box \hat{h}]
               = \frac{1}{2}(\Box \hat{h})\hat{M}(\Box \hat{h}) 
               + \frac{1}{2}\hat{h}\hat{N}(\Box \hat{h}) 
               +\hat{T}\hat{h}
               \quad .
\end{equation}

\bigskip

The particle interpretation of  (35) is now the central question. On one hand we can start from 
the HD theory (34) and, after inverting the projectors,  obtain the quartic propagator
\begin{eqnarray}
\Delta^{HD}[ \hat{h}] & = & \frac{2}{(c\Box+a) \Box}P^{(2)}
 +\frac{27}{4}\frac{ \lambda^2}{(\lambda -1)^2 [2(3b+c) \Box-a] \Box}P^{(W)} 
 \\
            &   & \mbox{} -\frac{2}{(\xi_1 \Box+\xi_3)\Box}P^{(1)}
    -\frac{27}{4}\frac{ \lambda^2}{(\lambda-1)^4[(\xi_1 -\xi_2 )
    \Box+\xi_3 ]\Box} P^{(S)}\nn\quad .
\end{eqnarray}
On the other hand, the quadratic propagators arising from the LD theory 
(analogous of(25)) for the new {\it hat} fields  are
\begin{eqnarray}
\Delta^{LD}[ \tilde{ \hat{h} } ] & = &  
                \frac{2}{a\Box} P^{(2)}
                -\frac{27}{4}\frac{ \lambda^2}{(\lambda -1)^2 a\Box} P^{(W)}
                \nonumber\\
&   & \mbox{}   -\frac{2}{\xi_3 \Box}P^{(1)}
                -\frac{27}{4}\frac{ \lambda^2}{(\lambda-1)^4 \xi_3 \Box} P^{(S)}\quad ,
      \\
\Delta^{LD}[ \tilde{ \hat{\pi} } ] & = & 
                -\frac{2c}{a(c\Box+a)}P^{(2)}
                +\frac{27}{4}
                \frac{\lambda^2}{(\lambda -1)^2} 
                \frac{2(3b+c)}{a[2(3b+c)\Box-a]}
                                                P^{(W)}
\nonumber\\
&   & \mbox{}   +\frac{2}{\xi_{3}} 
                        \frac{\xi_{1}}{(\xi_{1}\Box +\xi_{3})}P^{(1)}
                +\frac{27}{4}\frac{\lambda^2}{(\lambda -1)^4 \xi_{3}} 
                \frac{(\xi_{1} - \xi_{2})} 
                {[(\xi_{1} -\xi_{2} )\Box+\xi_{3} ]} 
                                P^{(S)}\nn\quad .
\end{eqnarray}
As expected, the LD quadratic propagators sum up to give the HD quartic one, namely
\begin{equation}
\Delta^{HD}[\hat{h}]=\Delta^{LD}[\tilde{ \hat{h}}]+\Delta^{LD}[\tilde{ \hat{\pi}}]
\end{equation}
Notice that if we had not performed the $Q$--transformation, the propagators would have been
\begin{eqnarray}
\Delta^{HD}[h]&=&Q\Delta^{HD}[\hat{h}]Q \nonumber  \\
\Delta^{LD}[\tilde{h}]&=&Q\Delta^{LD}[\tilde{\hat{h}} ]Q    \\
\Delta^{LD}[\tilde{\pi}]&=&Q\Delta^{LD}[\tilde{\hat{\pi}}]Q   \quad ,          \nonumber
\end{eqnarray}
with   the mixing $P^{\{SW\}}$ occurring in them.

        The DOF counting can be readily done on (37). Since we are dealing with a properly gauge 
fixed four-derivative theory, all the fields in $\hat {h}_{\mu\nu}$ do propagate and therefore we 
have a total of $20$ DOF ($10$ massless and $10$ massive). According to the dimensionality of the 
respective spin subspaces, they are distributed as 5, 3, 1 and 1 DOF for the spin-2, 1, 
$0_{S}$ and $0_{W}$ respectively, which sum up 10 DOF for the massless fields,
and the same for the massive ones. In the (massless) $\tilde{h}$-sector, 
the spin-2 space contains the two DOF of the graviton plus three gauge
DOF, and the remaining five DOF are also gauge ones. The $\tilde{\pi}$-sector
describes the five DOF of a spin-2 poltergeist with (squared) mass
$a/c$ [13] [17], one physical scalar DOF with  mass $-a/2(3b+c)$, 
three third ghosts with gauge-dependent masses $\xi_{3}/\xi_{1}$ and
one third ghost with mass $\xi_{3}/(\xi_{1}-\xi_{2})\;$.
\bigskip

In the absence of gauge fixing  only the projectors $P^{(2)}$ and $P^{(S)}$ are involved, and 
$P^{inv}$ in (7) can be inverted only in the restricted spin subspace $2\oplus S$. In principle, 
in that case one is left with eight DOF in the LD theory, namely, the massless graviton, the 
massive spin-2 poltergeist and the physical scalar. This is generally so as long as critical 
relationships between $a$, $b$ and $c$ are avoided [10] [13] that make the order-reducing 
procedure singular. In those cases some DOF may collapse and the theory may have  fewer than 
eight DOF and/or larger symmetries. Fast DOF-counting recipes for gauge theories can be found in 
[18]. In ordinary two-derivative gravity, each of the four gauge-group local parameters of Diff-
invariance accounts  for the  killing of {\it two} DOF, leaving the two DOF of the graviton out 
of the ten DOF of $h_{\mu\nu}\,$. In four-derivative gravity each gauge-group parameter instead 
kills {\it three} DOF so that the initial twenty DOF reduce to the eight DOF quoted above. The 
mechanism is well illustrated in four-derivative QED [15], where one initially has eight DOF and 
the gauge invariance suppress three of them, leaving one photon and one massive spin-1 
poltergeist.

\bigskip

As detailed in Appendix II, the free part of the $Q$-transformed LD theory can be made local for 
a particular choice of gauge parameters. However, using  $Q(\lambda)$, for $\lambda=b/(4b+c)$, 
moves the non-locality to the  source term, namely to the interactions. This can also be avoided 
by requiring that $\lambda=-2\,$, in which case $Q$ becomes a numerical matrix, but this gives 
rise to a condition on the parameters $b$ and $c$ of the starting Diff-invariant theory. Leaving 
aside the interpretation of such a restriction, this does not mean that we had anyway obtained a 
sum of independent Lagrangian theories for each (massless and massive, spin 2, 1, $0_{S}$ and 
$0_{W}$) particle, notwithstanding the fact that  the spin subspaces appear well separated. This 
is not posible, as illustrated for instance by the  fact [19] that there is no second-order 
tensor local theory for spin-1 fields.

\section{Faddeev-Popov compensating terms}

As usual, the gauge fixing term (3) together with the compensating (HD)
Faddeev-Popov  Lagrangian can be expressed as a coboundary in the 
BRST cohomology, namely
\begin{equation}
{\cal L}_{g}+{\cal L}_{FP}=-s
 \left[\bar{C}^{\alpha}{\cal G}_{\alpha \beta} \chi^{\beta} [h]
 + \frac{1}{2}\bar{C}^{\alpha}{\cal G}_{\alpha \beta}{\cal B}^{\beta} \right] 
 \quad ,
\end{equation}
where $\bar{C}$ are  FP fermion ghosts and ${\cal B}$ is an auxiliary commuting field.
\vfill
\eject

In order to study the propagators of the new fields $\bar{C}$ and $\cal B$, 
it suffices to consider the linearized objects
\begin{eqnarray}
\chi^{\beta}[h]&=&\chi^{\beta \mu \nu}h_{\mu \nu}\equiv
(\eta^{\beta \mu}\partial^{\nu}
-\lambda \eta^{\mu \nu}\partial^{\beta})h_{\mu \nu}\nn\\
{\cal G}^{(h)}_{\alpha \beta}&\equiv&\xi_{1}\eta_{\alpha \beta}\square 
                -\xi_{2}\partial_{\alpha}\partial_{\beta }
                +\xi_{3}\eta_{\alpha \beta} \\
        &=&(\xi_{1}\square +\xi_{3})\theta_{\alpha \beta}
+\left[ (\xi_{1}-\xi_{2})\square +\xi_{3} \right]\omega_{\alpha \beta}
\quad ,\nn
\end{eqnarray}
and the BRST symmetry given by the (linearized) Slavnov transformations
\begin{equation}
\begin{array}{l}
s h_{\mu \nu}={\cal D}_{\mu \nu ,\alpha}C^{\alpha}  
\\
s C^{\alpha}=0       
\\
s \bar{C}^{\alpha}= {\cal B}^{\alpha}  
\\
s {\cal B}^{\alpha}=0     \quad ,
\end{array}
\end{equation}
where 
\begin{equation}
{\cal D}_{\mu \nu ,\beta}=\eta_{\mu \beta}\partial_{\nu}+\eta_{\nu \beta}
\partial_{\mu}
\end{equation}
is the gauge symmetry generator. With the diagonalization
\begin{equation}
{\cal B}^{\alpha}=B^{\alpha}-\chi^{\alpha}[h]
\end{equation}
the linearized (40) becomes
\begin{equation}
{\cal L}^{(2)}_{g} +{\cal L}^{HD}_{FP}=
 \frac{1}{2}\chi^{\alpha}[h]{\cal G}^{(h)}_{\alpha \beta}\chi^{\beta}[h]
 +\bar{C}^{\alpha}{\cal G}_{\alpha \gamma}\chi^{\gamma \mu \nu}
 {\cal D}_{\mu \nu , \beta}C^{\beta} 
 -\frac{1}{2}B^{\alpha}{\cal G}^{(h)}_{\alpha \beta}B^{\beta}\quad , 
\end{equation}
and one has
\begin{equation}
\begin{array}{l}
s h_{\mu \nu}={\cal D}_{\mu \nu ,\alpha}C^{\alpha}  
\\
s C^{\alpha}=0       
\\
s \bar{C}^{\alpha}=B^{\alpha}-\chi^{\alpha}[h]  
\\
s  B^{\alpha}= \chi^{\alpha\mu\nu}
{\cal D}_{\mu \nu , \beta} C^{\beta}\quad .
\end{array}
\end{equation}
Of course, they reflect the trivially Abelian gauge symmetry group $G$ to which the infinitesimal 
diffeomorphisms reduce in  the linearized theory. In the complete non-polynomial theory there are 
couplings between $h_{\mu\nu}\,$, $C\,$, $\bar{C}$ and $ B\,$, and the non-Abelian symmetry 
yields a more complicated set of $S$-transformations in which, for instance $sC^{\alpha}\neq 0$. 
One should also notice that the $Q$-transformation used in Section 3 is ininfluent in the form of 
${\cal L}_{FP}$, as can be checked by computing it with the corresponding operators $\hat{{\cal 
D}}=Q^{-1}{\cal D}$ and $\hat{\chi}^{\alpha\mu\nu}=({\chi Q})^{\alpha \mu\nu}$.

\bigskip

\noindent{T}he fermionic sector of the FP Lagrangian above, namely 
\begin{equation}
{\cal L}^{HD}[\bar{C}C]
=\bar{C}^{\alpha} \left[ (\xi_{1} \Box +\xi_{3} )\Box \theta_{\alpha \beta}
+2(1-\lambda)((\xi_{1}-\xi_{2})\Box +\xi_{3})\Box \omega_{\alpha \beta} 
\right]
C^{\beta}
\end{equation}
is higher-derivative whereas, in constrast with ordinary second-order theories, now the auxiliary 
bosonic field $B$ {\it does propagate} according to the 
Lagrangian
\begin{equation}
{\cal L}[B]=-\frac{1}{2}B^{\alpha}\left[
(\xi_{1}\square +\xi_{3})\theta_{\alpha \beta}
+\left[ (\xi_{1}-\xi_{2})\square +\xi_{3} \right]\omega_{\alpha \beta}
\right]
B^{\beta}
\end{equation}
which is already LD and always local.
We can also perform an order-reduction of ${\cal L}^{HD}[\bar{C} C]\;$ (Appendix III), yielding
\begin{eqnarray}
{\cal L}^{LD}[\bar{E} E \bar{F} F] &=&
\bar{E}^{\alpha} 
\left( 
\xi_{3} \theta_{\alpha \beta} + 2(1-\lambda)\xi_{3} \omega_{\alpha \beta} 
\right) 
\Box E^{\beta} \\ \nonumber
&&\mbox{}-\bar{F}^{\alpha}  
 \left( \frac{\xi_{3}}{\xi_{1}}(\xi_{1}\Box + \xi_{3} )
\theta_{\alpha \beta}
+\frac{ 2(1-\lambda)\xi_{3}  }{\xi_{1}-\xi_{2}} 
 ((\xi_{1}-\xi_{2})\Box+\xi_{3})
\omega_{\alpha \beta} \right) F^{\beta}\quad ,
\end{eqnarray}
where $\, E^{\alpha}+F^{\alpha}=C^{\alpha}\,$ and 
$\,\bar{E}^{\alpha}+\bar{F}^{\alpha}=\bar{C}^{\alpha}\,$. The Lagrangian (49) is local for the same choice (see equations (102)-(104) in Appendix II)
of gauge  parameters that makes  ${\cal L}_{LD}$ in (25) local.
        
        From (48) one directly reads
\begin{equation}
\Delta[B]=
\frac{\theta}{\xi_{1}\square +\xi_{3}} 
        +\frac{\omega}{(\xi_{1}-\xi_{2})\square +\xi_{3}}
\quad ,
\end{equation}
whereas the HD (oriented) propagator 
\begin{eqnarray}
\Delta^{HD}[\bar{C} C]&=&
\frac{\theta}{(\xi_{1}\square +\xi_{3})\square}
+\frac{\omega}{2(1-\lambda)}\frac{1}{((\xi_{1}-\xi_{2})\square +\xi_{3})
\square}
\\
&=&\frac{\theta}{\xi_{3}} \left( \frac{1}{\square}
-\frac{\xi_{1}}{\xi_{1}\square +\xi_{3}}\right)
+\frac{\omega}{2(1-\lambda)\xi_{3}}
\left(\frac{1}{\square}
-\frac{ \xi_{1}-\xi_{2} }{(\xi_{1}-\xi_{2})\square+\xi_{3}}
\right)\nonumber
\end{eqnarray}
obtained from (47) splits into the (also oriented) LD propagators
\begin{eqnarray}
\Delta[\bar{E}E ]&=&
\frac{\theta}{\xi_{3}\Box}+\frac{\omega}
{2(1-\lambda )\xi_{3}\Box}\\
\Delta[\bar{F}F ] &=&
-\frac{\xi_{1}}{\xi_{3}(\xi_{1}\Box+\xi_{3})}\theta
-\frac{(\xi_{1}-\xi_{2})}{2(1-\lambda)\xi_{3}
((\xi_{1}-\xi_{2})\Box+\xi_{3})}\omega\quad ,
\end{eqnarray}        
that can be derived  from (49) as well.

In (52) one counts four FP {\it fermion} ghosts $E$ and four $\bar{E}$ with 
massless poles, giving eight {\it negative} loop
contibutions that compensate for the eight massless gauge ghosts quoted 
in  Section 3. The compensation of the third ghosts contains 
non trivial features which are characteristic to the HD theories. 
From (53) one 
has that the FP {\it fermion} ghosts $F$ and $\bar{F}$ give six 
{\it negative } loop  contributions with propagator poles at $\xi_{3}/\xi_{1}$ and 
two at $\xi_{3}/(\xi_{1}-\xi_{2})$. This over-compensates the 
(three plus one) third ghosts. Here is where
the new {\it boson}  FP ghosts $B$, 
propagating with (50), 
come to the rescue: they give three {\it positive } contributions with
poles at $\xi_{3}/\xi_{1}$ and one  at $\xi_{3}/(\xi_{1}-\xi_{2})$,
thus providing the complete cancellation 
of ghost loop contributions.

This matching of the ghost masses is a consequence of the interplay of the order-reducing and 
BRST procedures. The master relationship is 
\begin{equation}
{{\cal G}^{(h)}}^{-1}={{\cal G}^{(\tilde{h})}}^{-1}-{{\cal G}^{(\tilde{\pi})}}^{-1}\quad ,
\end{equation}
where the massive poles of the third ghosts are displayed by ${{\cal G}^{(h)}}^{-1}$ and 
${{\cal G}^{(\tilde{\pi})}}^{-1}$, the latter also having massless zero-modes.  To see it we find 
useful defining the differential operator
\begin{equation}
Z^{\alpha}_{\beta}\equiv \chi^{\alpha\mu\nu}{\cal D}_{\mu\nu , \beta} = 
\Box\left[\theta^{\alpha}_{\beta}+2(1-\lambda)\omega^{\alpha}_{\beta}\right]
\quad ,
\end{equation}
and the differential kernels
\begin{equation}
K^{(i)}_{\alpha\beta}\equiv {\cal 
G}^{(i)}_{\alpha\gamma}Z^{\gamma}_{\beta}\quad\quad\quad(i=h,\tilde{h},\tilde{\pi})
\end{equation}
occurring in the FP Lagrangians above and worked out in (47) and (49).
The poles of the (massless) gauge ghosts lie in $Z^{-1}$ whereas the operator ${{\cal 
G}^{(\tilde{h})}}^{-1}$ has no poles. From (54) and (56) it follows that 
\begin{equation}
{K^{(h)}}^{-1}={K^{(\tilde{h})}}^{-1}-{K^{(\tilde{\pi})}}^{-1}
\end{equation}
which also reads $ \,\Delta^{HD}[\bar{C}C]=\Delta[\bar{E}E]+\Delta[\bar{F}F]\,$. Thus the $E$-
fields and the $F$-fields inherit the massless and the massive poles respectively. On the other 
hand $\,\Delta[B]=-{{\cal G}^{(h)}}^{-1}\,$,$\,$ so that also the boson $B$-fields share the same 
massive poles.

\bigskip        
\bigskip        

        The symmetries of $\,{\cal L}_{g}^{LD}+{\cal L}_{FP}^{LD}\,$ are
not trivial either. The symmetry of the (invariant part of the) HD theory under the group $G$ of 
the gauge variations
\begin{equation}
\delta h_{\mu \nu}={\cal D}_{\mu \nu ,\alpha}\varepsilon^{\alpha}\quad ,
\end{equation}
is inherited by the LD theory via (17) and (24) with the variations
\begin{eqnarray}
\delta\tilde{h}_{\mu \nu}&=&
               \left[ \bar{\eta}^{\rho \sigma}_{\mu \nu}
               +\Box(N^{-1}M)^{\rho \sigma}_{\mu \nu} \right]
{\cal D}_{\rho \sigma , \alpha} \varepsilon^{\alpha}
\nonumber \\
&=&
\left[ \frac{\xi_{1}\Box +\xi_{3}}{\xi_{3}} P^{(1) \rho \sigma}_{\mu \nu}
+\frac{(\xi_{1}-\xi_{2})\Box+\xi_{3}}{\xi_{3}} P^{(W) \rho \sigma}_{\mu \nu}          
\right] {\cal D}_{\rho \sigma , \alpha} \varepsilon^{\alpha} \quad , \\
\delta\tilde{\pi}_{\mu \nu} &=&
-\left[\Box(N^{-1}M)^{\rho \sigma}_{\mu \nu} \right]
{\cal D}_{\rho \sigma , \alpha} \varepsilon^{\alpha}
\nonumber \\
&=&
-\left[ \frac{\xi_{1}\Box }{\xi_{3}} P^{(1) \rho \sigma}_{\mu \nu}
+\frac{(\xi_{1}-\xi_{2})\Box}{\xi_{3}} P^{(W) \rho \sigma}_{\mu \nu}          
\right] {\cal D}_{\rho \sigma , \alpha} \varepsilon^{\alpha}\quad ,
\nonumber
\end{eqnarray}
both depending on the same four gauge-group parameters $\varepsilon^{\alpha}(x)$.
One may check that $\delta \tilde{h}_{\mu \nu}+\delta \tilde{\pi}_{\mu \nu}
=\delta h_{\mu \nu}$.
However, the {\it free} invariant part of the LD theory (25) 
exhibits a larger symmetry group, 
namely the fields $\tilde{h}$ and $\tilde{\pi}$  may be given
independent variations 
\begin{eqnarray}
\bar{\delta}\tilde{h}_{\mu \nu}&=&{\cal D}_{\mu \nu,\alpha}\varepsilon'^{\alpha}
\\
\bar{\bar{\delta}}\tilde{\pi}_{\mu \nu}&=&{\cal D}_{\mu \nu,\alpha}\varepsilon''^{\alpha} 
\quad ,
\end{eqnarray}
thus doubling the number of group parameters, 
with the  original symmetry as a diagonal-like subgroup  $G_{1}\subset G\times G\,$, which is 
isomorphic to $G$ [15]. 
One may then look at
\begin{equation}
{\cal L}_{g }^{LD}[\tilde{h}]\equiv \frac{1}{2}\chi[\tilde{h}]{\cal G}^{(\tilde{h})}
\chi[\tilde{h}]
\end{equation}
and 
\begin{equation}
{\cal L}_{g}^{LD} [\tilde{\pi}]
=-\frac{1}{2}\chi[\tilde{\pi}]{\cal G}^{(\tilde{\pi})} \chi[\tilde{\pi}]
\quad ,
\end{equation}
occurring in (27), as separate gauge fixings for the
symmetries (60) and (61) respectively, and wonder what happens with the
whole BRST scheme.

        The separate $S$-transformations would be
\begin{equation}
\begin{array}{lcccl}
s\tilde{h}_{\mu \nu}={\cal D}_{\mu \nu ,\alpha}E^{\alpha}  
&&&&      
s\tilde{\pi}_{\mu \nu}={\cal D}_{\mu \nu ,\alpha}F^{\alpha}
\\
sE^{\alpha}=0       
&&&&          
s F^{\alpha}=0  
\\
s \bar{E}^{\alpha}=B'^{\alpha}  -\chi^{\alpha}[\tilde{h}]
&&&&           
s \bar{F}^{\alpha}=B''^{\alpha}-\chi^{\alpha}[\tilde{\pi}]
\\
s B'^{\alpha}= \chi^{\alpha\mu\nu}     
{\cal D}_{\mu \nu ,\beta}E^{\beta}
&&&&         
s B''^{\alpha}=\chi^{\alpha\mu\nu}     
{\cal D}_{\mu \nu ,\beta}F^{\beta}\quad ,
\end{array}
\end{equation}
so we are led to write
\begin{eqnarray}
{\cal L}_{g}+{\cal L}^{\star}_{FP}&=&- s  \left[
\bar{E}^{\alpha}{\cal G}_{\alpha \beta}^{(\tilde{h})} \chi^{\beta}[\tilde{h}]
+\frac{1}{2}\bar{E}^{\alpha}{\cal G}_{\alpha \beta}^{(\tilde{h})} {\cal B}'^{\beta}
\right] \\
&&\mbox{}+  s   \left[
\bar{F}^{\alpha}{\cal G}_{\alpha \beta}^{(\tilde{\pi})} \chi^{\beta}[\tilde{\pi}]
+\frac{1}{2}\bar{F}^{\alpha}{\cal G}_{\alpha \beta}^{(\tilde{\pi})} {\cal B}''^{\beta}
\right] \nonumber \\
&=&  \mbox{}\frac{1}{2}\chi^{\alpha} [\tilde{h}]
                {\cal G}_{\alpha \beta}^{(\tilde{h})} 
        \chi^{\beta}[\tilde{h}]  
        -
     \mbox{}\frac{1}{2}\chi^{\alpha}[ \tilde{\pi} ]
                {\cal G}_{\alpha \beta}^{(\tilde{\pi})} 
        \chi^{\beta}[\tilde{\pi}]  \nonumber \\
& & \mbox{}     +\bar{E}^{\alpha}{\cal G}_{\alpha \rho}^{(\tilde{h})} 
      \chi^{\rho \mu \nu}
     {\cal D}_{\mu \nu , \beta} E^{\beta}
     -\bar{F}^{\alpha}{\cal G}_{\alpha \rho}^{(\tilde{\pi})} 
     \chi^{\rho\mu \nu}
     {\cal D}_{\mu \nu , \beta} F^{\beta}         \\
& & \mbox{}     -\frac{1}{2}B'^{\alpha}
     {\cal G}_{\alpha \beta}^{(\tilde{h})}  
     B'^{\beta}
     +\frac{1}{2}B''^{\alpha}
     {\cal G}_{\alpha \beta}^{(\tilde{\pi})}  
     B''^{\beta}\nn
\end{eqnarray}
Thus (49) agrees with the fermionic sector
of (66).

	Equations (64) define two cohomologies $\{\bar{V};s \}$ and $\{ \bar{\bar{V}};s\}$ with  
cohomological spaces $\bar{V}\equiv\{ \tilde{h},E,\bar{E},B' \}$  and $\bar{\bar{V}}\equiv\{ 
\tilde{\pi},F,\bar{F},B'' \}$ respectively, both being copies of the original $V\equiv 
\{h,C,\bar{C},B\}$ of (46) with boundary operator $s$. The polynomial (65) is then an exact 
cochain in the cohomology $\{\bar{V};s\}\oplus\{\bar{\bar{V}};s\}\equiv 
\{\bar{V}\oplus\bar{\bar{V}};s\oplus s \}$. 

The cohomology characterizing the HD theory appears as a subcohomology  $\{V_{1};s_{1}\}$  of the 
direct sum above. The subspace $V_{1}\subset\bar{V}\oplus \bar{\bar{V}}$ is defined by
\begin{equation}
\begin{array}{lcccl}
\tilde{h}_{\mu\nu}={\cal O}'^{\rho \sigma}_{\mu\nu }h_{\rho\sigma}
&&&&      
\tilde{\pi}_{\mu\nu}={\cal O}''^{\rho \sigma}_{\mu\nu }h_{\rho\sigma}
\\
E^{\alpha}={\cal O}'^{\alpha}_{\beta}C^{\beta}     
&&&&          
F^{\alpha}={\cal O}''^{\alpha}_{\beta}C^{\beta}  
\\
\bar{E}^{\alpha}={\cal O}'^{\alpha}_{\beta}\bar{C}^{\beta}      
&&&&          
\bar{F}^{\alpha}={\cal O}''^{\alpha}_{\beta}\bar{C}^{\beta}  
\\
 B'^{\alpha}= {\cal O}'^{\alpha}_{\beta}B^{\beta}&&&&         
B''^{\alpha}= {\cal O}'^{\alpha}_{\beta}B^{\beta}     
\quad ,
\end{array}
\end{equation}
where
\begin{eqnarray}
{\cal O}'^{\rho \sigma}_{\mu\nu }&\equiv& \bar{\eta}^{\rho \sigma}_{\mu \nu}
               +\Box(N^{-1}M)^{\rho \sigma}_{\mu \nu} \\
{\cal O}''^{\rho \sigma}_{\mu\nu }&\equiv&
               -\Box(N^{-1}M)^{\rho \sigma}_{\mu \nu} \\
{\cal O}'^{\alpha}_{\beta}&\equiv &\frac{\xi_{1}\Box +\xi_{3}}{\xi_{3}}\theta^{\alpha}_{\beta}
+\frac{(\xi_{1}-\xi_{2})\Box +\xi_{3}}{\xi_{3}} \omega^{\alpha}_{\beta}\\
{\cal O}''^{\alpha}_{\beta}&\equiv&
- \left( \frac{ \xi_{1} \Box }{\xi_{3}}
\theta^{\alpha}_{\beta}+\frac{(\xi_{1}-\xi_{2})\Box}{\xi_{3}}
\omega^{\alpha}_{\beta} \right)
\end{eqnarray}
are invertible linear operators satisfying ${\cal O}'+{\cal O}''=\delta$, and $s_{1}$ is the 
restriction to $V_{1}$ of $s\oplus s$. Then this subcohomology is nothing but the original one 
$\{V;s\}$ of the HD theory, since (67) defines an isomorfism $V\stackrel{\imath_{1}}{\rightarrow} 
V_{1}$  and $s_{1}$ becomes $s\,$, that is
\begin{eqnarray}
s\tilde{h}_{\mu \nu}+s\tilde{\pi}_{\mu\nu}&=&
sh_{\mu\nu}
\nonumber
\\
s E^{\alpha} + s F^{\alpha} &=&  sC^{\alpha}
\nonumber\\
s\bar{E}^{\alpha}+s\bar{F}^{\alpha} &=&
s\bar{C}^{\alpha}\\
sB'^{\alpha} +sB''^{\alpha}&=& sB^{\alpha}\quad ,
\nonumber
\end{eqnarray}
as a consequence  of (67) and (68)-(71).
In other words we have  $\imath_{1}^{-1}\circ s\oplus s \circ\imath_{1}=s$. Moreover, we recover 
the Lagrangian (48) for $B$, namely
\begin{eqnarray}
{\cal L}^{\star}[B'B'']&=&
-\frac{1}{2}B'^{\alpha}{\cal G}_{\alpha \beta}^{(\tilde{h})}B'^{\beta}
+\frac{1}{2}B''^{\alpha}{\cal G}_{\alpha \beta}^{(\tilde{\pi})}B''^{\beta} 
\nonumber \\
&=&
-\frac{1}{2}B^{\alpha}{\cal O}'^{\gamma}_{\alpha}
{\cal G}_{\gamma \rho}^{(\tilde{h})}{\cal O}'^{\rho}_{\beta}B^{\beta}
+\frac{1}{2}B^{\alpha}
{\cal O}''^{\gamma}_{\alpha}
{\cal G}_{\gamma \rho}^{(\tilde{\pi})}{\cal O}''^{\rho}_{\beta}B^{\beta}
\\
&=&-\frac{1}{2}B^{\alpha}{\cal G}_{\alpha \beta}^{(h)}B^{\beta}=
{\cal L}[B]\nn
\end{eqnarray}
The subgroup $G_{1}\subset G\times G$, associated to $\{ V_{1}; s_{1}\}$ and isomorphic to $G$, 
is obtained by taking the group parameters $\varepsilon'$ and $\varepsilon''$ as functions of  
four parameters $\varepsilon$ by means of the equations 
\begin{eqnarray}
{\cal D}_{\mu \nu ,\alpha}\varepsilon'^{\alpha}&=&
               {\cal O}'^{\rho\sigma}_{\mu\nu}
{\cal D}_{\rho \sigma , \alpha} \varepsilon^{\alpha}
 \\
{\cal D}_{\mu \nu ,\alpha}\varepsilon''^{\alpha} &=&
			{\cal O}''^{\rho\sigma}_{\mu\nu}
{\cal D}_{\rho \sigma , \alpha} \varepsilon^{\alpha}\quad .\nn
\end{eqnarray}
These are derived by imposing the relations stemming from (59) on the otherwise independent 
variations (60) and (61), and yield
\begin{eqnarray}
\varepsilon'^{\alpha}&=&
{\cal O}'^{\alpha}_{\beta}\varepsilon^{\beta}\\
\varepsilon''^{\alpha}&=&
 {\cal O}''^{\alpha}_{\beta} \varepsilon^{\beta} \quad ,
\end{eqnarray}
so that $\varepsilon=\varepsilon'+\varepsilon''\,$.
The subgroup $G_{1}$ is, by definition, a symmetry of the whole (non gauge-fixed) Lagrangian (and 
also separately of the interaction terms) since one has 
$\bar{\delta}\tilde{h}_{\mu\nu}+\bar{\bar{\delta}}\tilde{\pi}_{\mu\nu}=\delta h_{\mu\nu}\,$, 
whereas $G\times G$ is broken by  the interaction terms.

\section{Conclusion}
The interplay of gauge invariance and higher differential order in field theories gives rise to a 
remarkable diversity of particle-like states which are encripted in  the original field 
variables. In the four-derivative tensor theory of gravity here studied, the doubling of the 
initial conditions for the (fourth differential-order) equations of motion translates into a 
doubling of the effective number of particle-like DOF obeying second differential-order evolution 
equations. They describe physical (positive Fock-space norm) states together with an outburst of 
massless and massive ghostly states which are unphysical because of their negative norm and/or 
gauge dependence. Beyond the methodological interest, its analysis provides an enlarged context 
for the traditional gauge theories and BRST symmetries of physical relevance, which also 
enlightens the nature of some states already encountered in former classical works on higher-
derivative gravity.

Four-derivative gravity is particularly interesting to study as long as the emphasis 
traditionally given to its applications has overlooked many details of its theorical structure. 
Amongst the particle-like states of the gauge-fixed theory, there are physical ones (a massless 
graviton and one scalar, reminiscent of the Brans-Dicke field), a massive spin-2 gauge 
independent Weyl ghost (unphysical norm), and two families of gauge-dependent fields: the usual 
massless gauge ghosts and the novel massive third ghosts.
This elusive new breed of ghosts firstly arose in the exponentiation of the functional 
determinant of the differential operator ${\cal G}^{(h)}$.

In the presence of (generally HD) gauge fixing terms and of the corresponding compensating FP 
Lagrangian, the order-reducing procedure reveals remarkable features of the underlying BRST 
symmetry associated to the four-parameter gauge group $G$ of infinitesimal diffeomorphisms. In 
parallel with the doubling of the fields, there is a doubling of the gauge symmetry of the {\it 
free} part of the (second-order) LD equivalent theory. Out of this $G\times G$ larger symmetry, 
both the interaction terms and the consistency of the BRST algebra select a diagonal-like  
subgroup $G_{1}\,$, isomorphic to $G\,$, as the only symmetry of the complete LD theory, in 
agreement with the ocurrence of Diff-invariance as the only symmetry of the starting HD theory. 
However, restricting ourselves to the free LD theory and considering its $G\times G$ symmetry, 
the LD gauge-dependent terms can be viewed as separate gauge fixings for both group factors. The 
(gauge-dependent) unphysical  propagating DOF so introduced then appear as the respective gauge 
ghosts, which are massless for the first group factor and massive for the second one, thus giving  
further meaning to the famous third ghosts. Moreover, the separate symmetry of the the gauge-
independent part of the physical and poltergeist sectors of the LD theory illustrates also how 
their kinetic terms reproduce the structure of the Einstein's and Fierz-Pauli theory, thus 
describing  (massless and massive) spin-2 fields respectively. 

In correspondence with the appearance of a new class of massive gauge ghosts, the compensating FP 
Lagrangian also contains a greater number of propagating fiels. These come from the HD doubling 
of the FP anticommuting fermion fields and from the boson fields, which are just auxiliary 
decoupled artifacts in ordinary two-derivative theories and now propagate and couple to the 
gauge-independent fields. The negative loop contributions of the massive fermion FP fields yield 
twice the amount needed to compensate for the third ghost loops, and it is just the positive 
contributions of the boson  FP fields that provide  the exact balance. This striking compensation 
mechanism, peculiar to HD gauge theories and easy to extrapolate to higher than four-derivative 
theories, illustrates well the power and richness of the BRST procedure. Of course, checking the 
exact cancellation of ghost loop contributions would require considering the actual residues of 
the propagators and vertex couplings arising in the complete non-polynomial theory, a task which 
is beyond the purposes of this work.

A final comment on locality is in order. From a HD local theory, the order-reducing procedure 
leads to an equivalent two-derivative theory. For scalar theories, the LD counterpart is directly 
local [16]. In gauge theories of vector fields there is always a choice of the gauge fixing 
parameters for which it is also local [15]. For tensor fields, the example studied in this paper 
tells us that obtaining an equivalent LD theory with independent free Lagrangians for the 
different spin states is not compatible with locality, although one comes close to this goal by 
suitably picking the gauge parameters. This obstruction is related to the more complex structure 
of the constraints of the tensor field theories, like the one that prevents having a tensor local 
theory of second differential-orden for spin-1 fields.

\vfill
\eject

\section*{Appendix I}

We use the notations

\begin{eqnarray}
g_{\mu \nu} &=& \eta_{\mu \nu} + h_{\mu \nu} \nonumber\\
A^{\mu} & \equiv & \partial_{\nu}h^{\mu \nu} \nonumber\\
h & \equiv & h_{\mu}^{\mu} \nonumber\\
X_{(\mu \nu)} & \equiv & X_{\{\mu \nu\}}  \equiv X_{\mu \nu} + X_{\nu \mu} \quad 
,\nonumber\end{eqnarray}
and the Minkowsky metric is $ \eta_{\mu \nu} = diag(+---) $. 

\bigskip
        The spin projectors  are
\begin{eqnarray}
P^{(2)}_{\mu\nu,\rho\sigma}&=&\frac{1}{2}\theta_{\mu(\rho}\theta_{\nu\sigma)}
-\frac{1}{3}\theta_{\mu\nu}\theta_{\rho\sigma}\\
P^{(1)}_{\mu\nu,\rho\sigma}&=&\frac{1}{2}\theta_{\{\mu(\rho}\omega_{\nu\}\sigma)}\\
P^{(S)}_{\mu\nu,\rho\sigma}&=&\frac{1}{3}\theta_{\mu\nu}\theta_{\rho\sigma}\\
P^{(W)}_{\mu\nu,\rho\sigma}&=&\omega_{\mu\nu}\omega_{\rho\sigma}
\end{eqnarray}
They are symmetric under the interchanges
\begin{equation}
\mu\leftrightarrow\nu\quad,\quad 
\rho\leftrightarrow\sigma\quad,\quad\mu\nu\leftrightarrow\rho\sigma\quad ,
\end{equation}
idempotent, orthogonal to each other, and sum up to the identity operator in the space of 
symmetric two-tensors, namely
\begin{equation}\bar{\eta}_{\mu\nu,\rho\sigma}\equiv \frac{1}{2}\eta_{\mu(\rho}\eta_{\nu\sigma)}
=(P^{(2)}+P^{(1)}+P^{(S)}+P^{(W)})_{\mu\nu,\rho\sigma}\quad .
\end{equation}
These projectors are constructed using the transverse and longitudinal projectors for vectors 
fields
\begin{eqnarray}
\theta_{\mu\nu}&=&\eta_{\mu\nu}-\frac{\partial_{\mu}\partial_{\nu}}{\Box}\\
\omega_{\mu\nu}&=&\frac{\partial_{\mu}\partial_{\nu}}{\Box}\quad\quad\quad. 
\end{eqnarray}
We also use the transfer operators
\begin{eqnarray}
P^{(SW)}_{\mu\nu,\rho\sigma}&=&\theta_{\mu\nu}\omega_{\rho\sigma}\\
P^{(WS)}_{\mu\nu,\rho\sigma}&=&\omega_{\mu\nu}\theta_{\rho\sigma}
\end{eqnarray}
from which we define
\begin{equation}
P^{\{ SW \}}=P^{\{ WS\}}\equiv P^{(SW)}+P^{(WS)}\quad.
\end{equation}

\noindent{T}hey have non-zero products
\begin{eqnarray}
P^{(SW)}P^{(WS)}&=&3P^{(S)}\\
P^{(WS)}P^{(SW)}&=&3P^{(W)}\\
P^{(S)}P^{(SW)}&=&P^{(SW)}P^{(W)}=P^{(SW)}\\
P^{(W)}P^{(WS)}&=&P^{(WS)}P^{(S)}=P^{(WS)}\\
P^{\{ SW\} }P^{\{ SW\} }&=&3(P^{(S)}+P^{(W)})\\
P^{(S)}P^{\{ SW\} }&=&P^{\{ SW \}}P^{(W)}=P^{(SW)}\\
P^{(W)}P^{\{ SW\} }&=&P^{\{ SW\} }P^{(S)}=P^{(WS)}\quad.
\end{eqnarray}
We define also
\begin{equation}
\bar{\bar {\eta} }_{\mu\nu,\rho\sigma} \equiv 
\eta_{\mu\nu}\eta_{\rho\sigma}=3P^{(S)}+P^{(W)}+P^{\{SW\}}\quad.
\end{equation}
\bigskip

       Here we collect some formulae  which are useful for dealing with combinations of the 
operators above. Inverse:
\begin{eqnarray}
{\cal M}&=&\lambda_{2}P^{(2)} +       
\lambda_{1}P^{(1)}+\lambda_{S}P^{(S)}+\lambda_{W}P^{(W)}
+\lambda_{SW}P^{\{SW\}}\nonumber\\
{\cal M}^{-1}&=&\frac{1}{\lambda_{2}}P^{(2)} +       
        \frac{1}{\lambda_{1}}P^{(1)}
        +\frac{\lambda_{W}}{\lambda_{S}\lambda_{W}-3\lambda_{SW}^{2}} P^{(S)}
        +\frac{\lambda_{S}}{\lambda_{S}\lambda_{W}-3\lambda_{SW}^{2}}P^{(W)} 
        \nonumber\\
& & \mbox{} -\frac{\lambda_{SW}}{\lambda_{S}\lambda_{W}-3\lambda_{SW}^{2}}
                 P^{\{SW\}} \quad .
        \nonumber
\end{eqnarray}

\noindent{W}hen computing symmetric products like (26), operators in the subspace $S\oplus W$ of 
the form 
\begin{equation}
\Omega(\tau_{S},\tau_{W},\tau_{SW})=\tau_{S}P^{(S)}+\tau_{W}P^{(W)}+\tau_{SW}P^{\{SW\}}
\end{equation}
occur, for which one has the product law
\begin{eqnarray}
\Omega(\tau_{S},\tau_{W},\tau_{SW})\Omega(\lambda_{S},\lambda_{W},\lambda_{SW})
\Omega(\tau_{S},\tau_{W},\tau_{SW})&=&\\
\Omega(\tau_{S}^{2}\lambda_{S}+3\tau_{SW}^{2}\lambda_{W}+6\tau_{S}\tau_{SW}\lambda_{SW}&,&\nn\\
\tau_{W}^{2}\lambda_{W}+3\tau_{SW}^{2}\lambda_{S}\!\!\!&+&\!\!\!6\tau_{W}\tau_{SW}\lambda_{SW}
\,\,\; ,\nn\\
\tau_{S}\tau_{W}\lambda_{SW}\!\!\!&+&\!\!\!\tau_{S}\tau_{SW}\lambda_{S}+\tau_{SW}\tau_{W}\lambda_
{W}
+3\tau_{SW}^{2}\lambda_{SW})\nn
\end{eqnarray}
\bigskip

	A basis of zeroth differential-order local operators with the symmetries (81) is provided 
by $\bar{\eta}$ and $\bar{\bar{\eta}}$. Local second-order operators can be expanded in the basis
\begin{eqnarray}
{\cal C}_{1}\Box&:=& \left(\frac{1}{2} P^{(2)}-P^{(S)}\right) \square\nonumber\\
{\cal C}_{2}\Box&:=&\left( \frac{1}{2}P^{(1)}+3P^{(S)}\right) \square \nonumber\\  
{\cal C}_{3}\Box&:=&\left( P^{\{SW\}}+6P^{(S)}\right) \square  \\
{\cal C}_{4}\Box&:=&\left( P^{(W)}-3P^{(S)}\right) \square\nonumber
\end{eqnarray}
Thus, a general local LD operator has the form 
\begin{eqnarray}
\Omega^{LD}&=&\sum_{i=1}^{4} \alpha_{i} {\cal C}_{i} \Box
+ a_{1} \bar{\eta}+a_{2}\bar{\bar{\eta}}\quad .
   \end{eqnarray}

Einstein's (linearized) theory displays the operator ${\cal C}_{1}$. Fierz-Pauli's has the same 
kinetic term and a mass term built with $\bar{\eta}-\bar{\bar{\eta}}$. When using the field basis 
obtained by $Q(\lambda)$-transforming the theory, this kinetic term displays the operator
\begin{equation}
Q(\lambda) {\cal C}_{1} \Box Q(\lambda) =
        \left( \frac{1}{2} P^{(2)} 
        -\frac{4}{27} \frac{(\lambda -1)^{2}}{\lambda^{2}}P^{(W)}\right)\Box \quad
\end{equation}
For $\lambda=-2$ it becomes local again, namely
$\left( \frac{1}{2}P^{(2)}-\frac{1}{3}P^{(W)} \right)\Box
={\cal C}_{1}\Box -\frac{1}{3}{\cal C}_{3}\Box $ 
which describes (linearized) gravity as properly as the original operator ${\cal C}_{1}\Box$ did.

\bigskip
\bigskip

\section*{Appendix II}

The kinetic terms for $\tilde h$ and $\tilde \pi$ in (25) contain the operator $N\Box$
which is local for arbitrary gauge parameters. In fact one has that
\begin{equation}
N=a{\cal C}_{1}
-\xi_{3}{\cal C}_{2}
-\lambda(\lambda-1)\xi_{3}{\cal C}_{3}
-\xi_{3}(\lambda-1)^{2}{\cal C}_{4}
\end{equation}
However, the ``mass term''
$ -\frac{1}{2} \tilde{\pi} NM^{-1}N \tilde{\pi} $ is local only for a choice of gauge parameters 
obeying the conditions
\begin{eqnarray}
\xi_{1} &=& -c\frac{\xi_{3}^{2}}{a^{2}}\quad , \\
\frac{a^{2}}{2c}\frac{\xi_{1}-\xi_{2}}{\xi_{3}^{2}} &=& -\frac{3b+c}{4b+c}  
\quad , \\
\lambda &=& \frac{b}{4b+c}\quad ,
\end{eqnarray}
that are obtained by requiring $NM^{-1}N$ to be a linear combination of 
$\bar{\eta}$ and $\bar{\bar{\eta}}$. This leaves one of the parameters $\xi$
still arbitrary.
This same conditions make the fermion Lagrangian (48) local.

In view of the conditions above, a theory in which $4b+c=0$ does not have
a local LD equivalent. But this case was critical already for the complete Diff-invariant theory 
since it is not regular in $R$ and a general-covariant Legendre transform (see equation (8)
of [13]) cannot be performed. With gauge fixing terms and for the 
linearized field, the (just Lorentz-covariant) Legendre transform can
always be carried out (equation (17) is not singular for $4b+c=0$)
and we instead have non-locality of the LD theory.

\bigskip

When we consider the $Q$-transformed theory, the potentially non-local
operator is
\begin{eqnarray}
 \hat{N}\hat{M}^{-1}{\hat N}&=&\frac{a^2}{2c}P^{(2)}
 +\frac{4}{27}\frac{(\lambda -1)^2}{\lambda^{2}} \frac{a^{2}}{2(3b+c)}P^{(W)}
 \nonumber\\
&-&\frac{\xi_{3}^{2}}{2\xi_{1}}P^{(1)}
-\frac{4}{27}\frac{(\lambda-1)^{4}}{\lambda^{2}}
\frac{\xi_{3}^{2}}{\xi_{1}-\xi_{2}}P^{(S)}\quad .
\end{eqnarray}
As explained before, we must take $\lambda=-2$ in order to keep
the locality of the source term. In that case $\hat{N}\Box$ remains local.
Requiring locality for $\hat{N}\hat{M}^{-1}\hat{N}$ leads to 
\begin{eqnarray}
\xi_{1}=-\frac{1}{5}\xi_{2}\\
\xi_{3}^{2}=\frac{a^{2}}{5c}\xi_{2}\\
c=-\frac{9}{2}b
\end{eqnarray}
These conditions are the same found before, but now (104) yields a 
condition, namely (108), on the parameters of the original gauge-invariant theory.

\bigskip
\bigskip

\section*{Appendix III}
We briefly outline the order-reduction of the higher-derivative FP Lagrangian for anticommuting 
fields
\begin{equation}
{\cal L}_{HD}= \bar{C}_{\mu}\, 
        \left(
                \square(a_1\square +b_1)\,\theta^{\mu \nu}
                +\square(a_2\square +b_2)\,\omega^{\mu \nu}
                \right)\,C_{\nu} +\bar{\zeta}^{\mu}C_{\mu}
                +\bar{C}_{\mu}\zeta^{\mu}
\end{equation}
where $\zeta$ and $\bar{\zeta}$ are external sources which are also anticommuting. Dropping total 
spacetime derivatives, conjugate momenta may be defined as the left  derivatives
\begin{eqnarray}
{\cal P}^{\mu}&=&\frac{\partial^{L} {\cal L}_{HD}}
                {\partial \,\square \,\bar{C}_{\mu}}=
                {\cal M}^{\mu \nu} \square\,C_{\nu} 
                +\frac{1}{2}{\cal N}^{\mu \nu} C_{\nu}\\
\bar{\cal P}^{\mu}&=&\frac{\partial^{L} {\cal L}_{HD}}
                {\partial \,\square\, C_{\mu}}=
                -{\cal M}^{\mu \nu} \square \,\bar{C}_{\nu} 
                -\frac{1}{2}{\cal N}^{\mu \nu} \bar{C}_{\nu}
\end{eqnarray}
where
\begin{equation}
\begin{array}{lr}
{\cal M}\equiv a_1\theta +a_2\omega \quad, & {\cal N}\equiv b_1\theta +b_2\omega\quad ,
\end{array}
\end{equation}
from which
\begin{eqnarray}
\square\, C_{\mu}& =&{\cal M}^{-1}_{\mu \nu}\left(
        {\cal P}^{\nu}-\frac{1}{2}{\cal N}^{\nu \rho} C_{\rho} \right)\\
\square \,{\bar C}_{\mu}& =&-{\cal M}^{-1}_{\mu \nu}\left(
        \bar{\cal P}^{\nu}+\frac{1}{2}{\cal N}^{\nu \rho}\bar{C}_{\rho} 
        \right)
\end{eqnarray}
	Then the ``Hamiltonian'' is
\begin{eqnarray}
{\cal H} &\equiv&(\Box \bar{C}){\cal P}+(\Box C) {\cal P}-{\cal L}\nonumber\\
&=& -\left(\bar{\cal P} +\frac{1}{2}{\cal N}\bar{C}\right)
              {\cal M}^{-1}
            \left({\cal P} -\frac{1}{2}{\cal N}C \right) 
                 -\bar{\zeta}^{\mu}C_{\mu}
                -\bar{C}_{\mu}\zeta^{\mu}
\end{eqnarray}
With the field redefinition 
\begin{equation}
\begin{array}{ll}
C=E+F&\bar{C}=\bar{E}+\bar{F}\\
{\cal P}=\frac{1}{2}{\cal N}\left( E-F\right)&\bar{{\cal P}}=\frac{1}{2}{\cal N}\left(\bar{F}-
\bar{E}\right)
\end{array}
\end{equation}
the Helmholtz Lagrangian
\begin{equation}
{\cal L}_{H}\equiv(\square \,C)\bar{\cal P} +(\square\, \bar{C}){\cal P} -{\cal H}
\end{equation}
becomes
\begin{eqnarray}
{\cal L}_{LD}&=& \bar{E}{\cal N} \square\, E
        -\bar{F} 
                \left( {\cal N}\square +{\cal N}{\cal M}^{-1}{\cal N} \right) 
                 F \nonumber\\
& & \mbox{} +\bar{\zeta}( E + F )
        +(\bar{E}+\bar{F})\zeta\quad .
\end{eqnarray}

\vfill
\eject

\section*{References}

\noindent [1] D.J.Gross and E.Witten, 
                                 {\it Nucl.Phys.}{\bf B277}(1986)1. 

              R.R.Metsaev and A.A.Tseytlin, 
                               {\it Phys.Lett.}{\bf B185}(1987)52 .
 
              M.C.Bento and O.Bertolami, 
                               {\it Phys.Lett.}{\bf B228}(1989)348.      
 
\noindent [2] N.D.Birrell and P.C.W.Davies, {\it Quantum Fields in
                 Curved Space},
           
                 Cambridge Univ.Press(1982).

\noindent [3] K.S.Stelle, {\it Phys.Rev.}{\bf D16}(1977)953.

\noindent [4] J.Julve and M.Tonin, 
                                {\it Nuovo Cim.}{\bf B46}(1978)137.

\noindent [5] E.S.Fradkin and A.A.Tseytlin, 
                               {\it Nucl.Phys.}{\bf B201}(1982)469.

\noindent [6] N.H.Barth and S.M.Christensen,
                                  {\it Phys.Rev.}{\bf D28}(1983)1876.

\noindent [7] I.B.Avramidy and A.O.Barvinsky, 
                                {\it Phys.Lett.}{\bf 159}(1985)269.

\noindent [8] I.Antoniadis and  E.T.Tomboulis,
					{\it Phys.Rev.}{\bf D33}(1986)2756.

\noindent [9] T.Goldman, J.P\'erez-Mercader, F.Cooper and M.M.Nieto,
                                  {\it Phys.Lett.}{\bf 281}(1992)219.      
            
              E.Elizalde, S.D.Odintsov and A.Romeo,
                                  {\it Phys.Rev.}{\bf D51}(1995)1680.

\noindent [10] I.L.Buchbinder, S.D.Odintsov and I.L.Shapiro, 

         {\it Effective Action in Quantum Gravity},                                                                                                                                                                                

          (IOP, Bristol and Philadelphia, 1992).

\noindent [11] S.Deser and  D.G.Boulware,
                                 {\it Phys.Rev.}{\bf D6}(1972)3368.

               B.Whitt          {\it Phys.Lett.}{\bf B145}(1984)176.
               
               J.D.Barrow and S.Cotsakis,
                                {\it Phys.Lett.}{\bf B214}(1988)515.
               
               G.Gibbons,    {\it Phys.Rev.Lett.}{\bf 64}(1990)123 .               

\noindent [12] M.Ferraris and J.Kijowski, 
                                {\it Gen.Rel.Grav.}{\bf 14}(1982)165.

              A.Jakubiec and J.Kijowski, 
                                  {\it Phys.Rev.}{\bf D37}(1988)1406.
                                                            
              G.Magnano, M.Ferraris and M.Francaviglia,
                               {\it Gen.Rel.Grav.}{\bf 19}(1987)465;
                                  
                           {\it J.Math.Phys.}{\bf 31}(1990)378;
                           {\it Class.Quantum.Grav.}{\bf 7}(1990)557.

\noindent [13] J.C.Alonso, F.Barbero, J.Julve and A.Tiemblo,
                           {\it Class.Quantum Grav.}{\bf11}(1994)865.

\noindent [14] A.Hindawi, B.A.Ovrut and D.Waldram,
                         {\it Phys.Rev.}{\bf D53}(1996)5583.

\noindent [15] A.Bartoli and J.Julve, 
                               {\it Nucl.Phys.}{\bf B425}(1994)277.

\noindent [16] F.J.de Urries and J.Julve,
                       {\it J.Phys.A:Math.Gen.}{\bf 31}(1998)6949.

\noindent [17] M.Fierz and W.Pauli, 
                                {\it Proc.R.Soc.}{\bf A173}(1939)211. 

\noindent [18] M.Henneaux, C.Teitelboim and J.Zanelli,
					{\it Nucl. Phys.}{\bf B332}(1990)169.

\noindent [19] P. van Nieuwenhuizen, 
                                  {\it Nucl.Phys.}{\bf B60}(1973)478.

\end{document}